\documentclass[acmsmall,screen]{acmart}

\usepackage{pifont}
\usepackage{textcomp}
\usepackage{bbm}
\usepackage{listings}
\usepackage{makecell}
\usepackage{multirow}
\usepackage{tcolorbox}

\newcommand{\tool}[1]{{ACECode}}

\AtBeginDocument{%
  }

\setcopyright{acmlicensed}
\copyrightyear{2024}
\acmYear{2024}
\begin{document}

%%
%% The "title" command has an optional parameter,
%% allowing the author to define a "short title" to be used in page headers.
\title{ACECode: A Reinforcement Learning Framework for Aligning Code Efficiency and Correctness in Code Language Models}

%%
%% The "author" command and its associated commands are used to define
%% the authors and their affiliations.
%% Of note is the shared affiliation of the first two authors, and the
%% "authornote" and "authornotemark" commands
%% used to denote shared contribution to the research.
\author{Chengran Yang}
\email{cryang@smu.edu.sg}
\affiliation{%
  \institution{Singapore Management University}
  \country{Singapore}
}

\author{Hong Jin Kang}
\email{hjkang@smu.edu.sg}
\affiliation{%
  \institution{Singapore Management University}
  \country{Singapore}
}

\author{Jieke Shi}
\email{jiekeshi@smu.edu.sg}
\affiliation{%
  \institution{Singapore Management University}
  \country{Singapore}
}

% \author{Zhensu Sun}
% \affiliation{%
%  \institution{Rajiv Gandhi University}
%  \city{Doimukh}
%  \state{Arunachal Pradesh}
%  \country{India}}

\author{David Lo}
\email{davidlo@smu.edu.sg}
\affiliation{%
  \institution{Singapore Management University}
  \country{Singapore}
}
%%
%% By default, the full list of authors will be used in the page
%% headers. Often, this list is too long, and will overlap
%% other information printed in the page headers. This command allows
%% the author to define a more concise list
%% of authors' names for this purpose.
% \renewcommand{\shortauthors}{Chengran et al.}

%%
%% The abstract is a short summary of the work to be presented in the
%% article.
\begin{abstract}

  CodeLLMs have demonstrated remarkable advancements in software engineering tasks. However, while these models can generate functionally correct code, they often produce code that is inefficient in terms of runtime. This inefficiency is particularly problematic in resource-constrained environments, impacting software performance and sustainability.

  Existing approaches for optimizing code efficiency for CodeLLMs like SOAP and PIE exhibit certain limitations. SOAP requires a compatible execution environment and predefined test cases for iterative code modification, while PIE focuses on instruction tuning, improving efficiency but compromising correctness. These shortcomings highlight the need for a fine-tuning framework that optimizes both efficiency and correctness without relying on predefined test cases or specific execution environments.

  To bridge this gap, we introduce ACECode, a reinforcement learning-based fine-tuning framework that aligns CodeLLMs with dual objectives of efficiency and correctness. ACECode combines three key steps: (1) generating code with an actor CodeLLM, (2) calculating a training-free reward signal derived from code execution feedback for each generated code, and (3) optimizing the CodeLLM via Proximal Policy Optimization (PPO) algorithm. This reward signal enables joint assessment of efficiency and correctness without manual labeling.

  We evaluate \tool{} by fine-tuning four SOTA (state-of-the-art) CodeLLMs and comparing their code with three baselines: original, instruction-tuned, and PIE-tuned CodeLLMs. Extensive experiment 
  results suggest that \tool{} significantly improves both the efficiency 
  and correctness of generated code against all baselines for all CodeLLMs. Specifically, CodeLLMs fine-tuned with \tool{} improve pass@1 by 1.84\% to 14.51\% and reduce runtime in 65\% to 72\% of cases compared to original CodeLLMs.

\end{abstract}

%%
%% The code below is generated by the tool at http://dl.acm.org/ccs.cfm.
%% Please copy and paste the code instead of the example below.
%%
  

% \received{9 December 2024}
% \received[revised]{12 March 2009}
% \received[accepted]{5 June 2009}


%%
%% This command processes the author and affiliation and title
%% information and builds the first part of the formatted document.
\maketitle

% \cran{Update the conference information}

\section{Introduction}
Large Language Models (LLMs), such as GPT~\cite{achiam2023gpt} and Qwen~\cite{bai2023qwen}, have demonstrated remarkable advancements across various domains including software engineering~\cite{wang2024software,fan2023large,hou2023large}. 
Specialized LLMs for code like Code Alpaca~\cite{codealpaca}, WizardCoder~\cite{luo2023wizardcoder}, Magicoder~\cite{wei2023magicoder}, and WaveCoder~\cite{yu2023wavecoder}, known as CodeLLMs, are built on general-purpose LLMs and fine-tuned with code datasets spanning multiple programming languages. These CodeLLMs demonstrate exceptional performance across a range of code generation tasks, such as vulnerability patch generation~\cite{pearce2023llmrepair} and code completion~\cite{guo2023codecompletion, izadi2024codecompletion}.

While current CodeLLMs are capable of generating functionally correct code,
this often comes at the expense of {\it code efficiency}. CodeLLM-generated code tends to be suboptimal in terms of runtime (i.e., the execution time for generated code), with prior studies~\cite{huang2024effibench, huang2024soap, shi2024efficientcode, niu2024evaluatingefficiency} showing that it can require 3-13$\times$ more runtime than human-written code.

The inefficiency of CodeLLM-generated code poses a significant challenge as CodeLLMs are increasingly integrated into real-world software development workflows. 
For instance, Google reports that CodeLLMs now generate over a quarter of new code in their products~\cite{forbesWritesOver}. 
As such, inefficient code generated by CodeLLMs can significantly hinders the performance of software products and reduces their competitiveness, particularly in resource-constrained scenarios like mobile devices, cloud servers, and IoT systems. 
On the other hand, optimizing the efficiency of CodeLLM-generated code can contribute to environmental sustainability by enabling software products to consume less energy and resources, ultimately lowering their carbon footprint~\cite{shi2024efficientcode}.

To date, few approaches have attempted to address this issue, including SOAP~\cite{huang2024soap} and PIE~\cite{shypula2023learning}. 
SOAP~\cite{huang2024soap} adopts a two-stage inference process, where CodeLLMs first generate code that is executed with test cases to collect runtime feedback. This feedback is then used to guide a second round of code generation aimed at reducing execution overhead.
However, SOAP requires a compatible execution environment for LLM-generated code and pre-defined test cases for each task, which significantly restricts its out-of-the-box usability. 
Moreover, the two-stage inference process doubles the inference time and introduces additional overhead during code execution

On the other hand, PIE~\cite{shypula2023learning} 
crafts a dataset with efficient code snippets and applies instruction tuning for CodeLLMs to optimize code efficiency.
While PIE optimizes code efficiency without relying on execution feedback, it applies instruction tuning with next-token prediction as its training objective. 
This approach primarily optimizes CodeLLMs as general language models, rather than addressing task-specific, multi-objective goals including code efficiency and correctness. Consequently, it leaves a gap in tailored fine-tuning methods specifically designed to optimize code efficiency.
Moreover, a recent empirical study~\cite{waghjale2024ecco} highlights a crucial drawback of existing approaches including PIE: while code efficiency improves with these methods, they often do so at the expense of functional correctness (referred to as correctness).
This drawback underscores the need for a fine-tuning framework specifically designed to optimize both code efficiency and correctness in CodeLLMs, eliminating the dependency on pre-defined test cases and a compatible execution environment during inference stage.

To address these limitations of existing approaches, we draw on techniques used for LLM alignment~\cite{ouyang2022training, wang2023aligningllm, shen2023aligningllm}—a methodology originally developed to tailor general LLM outputs to human preferences.
LLM alignment often applies reinforcement learning from human feedback (RLHF) as a fine-tuning approach~\cite{ouyang2022training, wang2023aligningllm, shen2023aligningllm}. 
In RLHF, the LLM generates responses evaluated by a reward model (referred to as a "rewarder"), which estimates how much a human might favor the responses.
The LLM’s parameters are then iteratively updated to maximize the reward signals with reinforcement learning algorithms, progressively aligning the model with human preferences.
% , followed by LLM parameter updating to maximiz the rewards. 
% This iterative process progressively aligns the LLM with human preferences over time.
In this work, we extend RLHF principles by tailoring specific LLM output, i.e., \textit{generated code}, to meet fine-grained code quality requirements, i.e., \textit{code efficiency and correctness}.

Applying RLHF to CodeLLMs to align the CodeLLM-generated code with correctness and efficiency requirements presents two primary challenges. First, RLHF typically requires a manually labeled dataset where LLM outputs are annotated with human preferences~\cite{ouyang2022training, dai2023safe, wang2023aligningllm, shen2023aligningllm} to train the reward model. Creating such a dataset is time-consuming and labor-intensive, requiring deep expertise in code algorithms and carrying the risk of subjective biases. The second challenge lies in representing both code efficiency and correctness within a rewarder, enabling the model to optimize for the dual objectives.

To address both challenges, we propose \tool{} (\textbf{A}ligning Code \textbf{C}orrectness and \textbf{E}fficiency for \textbf{Code}LLMs), a reinforcement learning-based fine-tuning framework incorporating a training-free rewarder derived from code execution. 
ACECode eliminating the need for manually labeled datasets to train a reward model.
Instead, it introduces a training-free, dual-objective rewarder that jointly evaluates code efficiency and correctness.
% In ACECode, we argue that training rewarders by human-labeled data is unnecessary; instead, our training-free rewarder assesses code correctness based on the execution status of code given the test cases and measures code efficiency by comparing its runtime to that of a human-written reference solution. 
% In ACECode, we argue that human-labeled datasets for training rewarders are unnecessary. Instead, the training-free rewarder 
Specifically, our rewarder assesses correctness based on the execution status of code against provided test cases and evaluates efficiency by comparing the runtime of generated code to that of a human-written reference solution.
Moreover, to represent efficiency and correctness within one rewarder, we design the rewarder as a step function that penalizes incorrect or inefficient code while rewarding code that achieves both attributes, guiding the model to optimize toward both goals jointly. With this reward feedback, the CodeLLM is then optimized using Proximal Policy Optimization (PPO)~\cite{schulman2017proximal} algorithm. To the best of our knowledge, this is the first RLHF method that improves both the efficiency and correctness of CodeLLM-generated code.

To evaluate the effectiveness of \tool{}, 
we conduct a series of experiments using an extended version of EffiBench~\cite{huang2024effibench}, a widely-used benchmark for code efficiency.
Specifically, to enable a comprehensive evaluation of efficiency and correctness, we augment each coding task in EffiBench, originally paired with one ground-truth solution, with 10 additional human-written ground-truth solutions.
% We extend EffiBench by augmenting each coding task, originally paired with one ground-truth solution, with 10 additional human-written ground-truth solutions. This extension allows for a more comprehensive evaluation of efficiency and correctness.
% we conduct a series of experiments on an extended version of commonly-used code efficiency benchmark, i.e., EffiBench~\cite{huang2024effibench}, with four state-of-the-art open-source CodeLLMs.
% We extend the each pair of code questions and one ground-truth solution in EffiBench with additional 10 ground-truth human-written solutions.
We assess the performance of \tool{} by fine-tuning four state-of-the-art open-source CodeLLMs and comparing their generated code with three baselines: original CodeLLMs, instruction-tuned CodeLLMs, and CodeLLMs fine-tuned with the state-of-the-art approach PIE~\cite{shypula2023learning}.
Extensive experiment results suggest that \tool{} significantly improves both the efficiency and correctness of generated code against all baselines for all CodeLLMs.
Specifically, CodeLLMs fine-tuned with \tool{} demonstrate improvements over the original CodeLLMs in code correctness ranging from 1.84\% to 14.51\% in terms of pass@1. 
Moreover, 65\% to 72\% of the code solutions generated by CodeLLMs fine-tuned with \tool{} exhibit reduced runtime compared to those produced by the original CodeLLMs.
Furthermore, ACECode outperforms instruction tuning and PIE in terms of code correctness
and efficiency, achieving improvements of up to 51.15\% in pass@1 and 23.18\% in average execution time
compared to instruction tuning, and up to 14.41\% in pass@1 and 11.45\% in average execution time
compared to PIE.

We summarize our contributions in this paper as follows:
\begin{itemize}
    \item \textbf{The first RLHF framework for optimizing dual attributes of LLM-generated code}: We introduce \tool{}, the first reinforcement learning-based fine-tuning framework for optimizing code efficiency and correctness.
    \item \textbf{Proposing a training-free rewarder for assessing code correctness and efficiency}: \tool{} provides a novel rewarder that jointly assesses code efficiency and correctness, derived directly from code execution feedback without requiring a manually labeled dataset.
    \item \textbf{Extending Existing Benchmark:} We extend existing benchmark EffiBench~\cite{huang2024effibench} by adding 10 additional human-written ground-truth solutions for each task, making the dataset more robust for evaluating efficiency and correctness improvements. 
    \item \textbf{Demonstrating the effectiveness of \tool{}}: We evaluate \tool{}'s effectiveness on the EffiBench benchmark with multiple state-of-the-art open-source CodeLLMs. Results show that \tool{} significantly enhances code efficiency and correctness, achieving improvements of up to 14.51\% in pass@1 rate and up to 10.86\% in average execution time against vanilla CodeLLMs. 
    % Our artifact is available at \url{https://github.com/autumn-city/Acecoder}.
\end{itemize}

The remaining part of this paper is organized as follows. Section 2 provides an overview of background concepts, related work, and formalizes the problem definition. Section 3 introduces ACECode, detailing its architecture and reward design. Section 4 presents the experimental setup, evaluation metrics, and implementation details. Section 5 discusses the results of our work. Section 6 discusses broader implications, multi-objective optimization, and threats to validity. Finally, Section 7 concludes the paper and outlines future work.

% \begin{figure}
%     \centering
%     \includegraphics[width=0.5\textwidth]{figures/overview.pdf}
%     \caption{The overview of \tool{}.}
%     \label{fig:overview}
% \end{figure}

\section{Background and Related Work}
\subsection{Related Work}
\subsubsection{LLMs for Code}
Recent research has focused on improving the capacity of LLMs to perform code-related tasks.
Gnerally, two lines of research have been proposed to enhance the performance of LLMs in code Generation: (1) pre-training LLM on large-scale code corpora, like AlphaCode~\cite{li2022alphacode}, StarCoder~\cite{lozhkov2024starcoder}, and CodeT5~\cite{wang2023codet5+}, and (2) fine-tuning foundamental general-purpose LLMs on code-related tasks or with code-related instruction datasets, like WizardCoder~\cite{luo2023wizardcoder}, Magicoder~\cite{wei2023magicoder}, and WaveCoder~\cite{yu2023wavecoder}.
Those LLMs specifically optimized for code-related tasks are referred to as CodeLLMs.
However, existing empirical studies~\cite{waghjale2024ecco, huang2024effibench} show that those CodeLLMs often overlook optimizing generated code's execution efficiency, resulting in 3 to 13$\times$ slower code than human-written peers. 

Limited works including SOAP~\cite{huang2024soap} and PIE~\cite{shypula2023learning} have attempted to optimize the execution efficiency of CodeLLM-generated code. 
SOAP~\cite{huang2024soap} leverages in-context learning with multiple iterations to enhance CodeLLM efficiency during inference. Specifically, SOAP operates within a code execution environment and iteratively prompts the CodeLLM to refine its generated code by providing execution feedback after each iteration.
However, SOAP requires a compatible execution environment and introduces additional overhead to model generation, which hinders its out-of-box usage.
PIE~\cite{shypula2023learning} improves code efficiency on C++ by creating a dataset of efficient code snippets and applying instruction-based tuning~\cite{waghjale2024ecco}, though it sacrifices code correctness.
% Unlike these approaches, our framework \tool{}, is execution enviroment dependency-free and requires no additional overhead at the inference stage, which simultaneously enhances both code correctness and efficiency in CodeLLMs.
In contrast, CodeLLMs fine-tuned by \tool{} are free from execution environment dependencies, incur no additional overhead during inference, and demonstrate improvements in both code correctness and efficiency, overcoming the limitations of previous approaches~\cite{huang2024soap, shypula2023learning}.

% While these approaches enhance CodeLLMs in generating functionally correct code, they often overlook the optimization of execution efficiency, an area critical for real-world deployment~\cite{huang2024effibench}.
% Our ACECode framework is adaptable and can be applied to both pre-trained and fine-tuned CodeLLMs to further improve the efficiency and correctness of the generated code.

\subsubsection{Evaluating CodeLLM for Code Generation}

Code generation, which involves generating code snippets or functions based on natural language prompts or docstrings, has received increasing attention in recent years~\cite{li2022alphacode,yu2023wavecoder,wei2023magicoder}. 
To evaluate LLM performance on this task, various benchmarks have been proposed, primarily focusing on the correctness of generated code across different scenarios.
For example, HumanEval assesses functional correctness by executing generated code on a set of test cases~\cite{chen2021humaneval}.
DS1000 evaluates code generation capabilities with a focus on third-party libraries within the data science domain~\cite{lai2023ds1000}.
EvoCodeBench measures code generation performance at the repository level~\cite{li2024repobench}.
Recently, EffiBench~\cite{huang2024effibench} and ECCO~\cite{waghjale2024ecco} are introduced to evaluate the efficiency of generated code. 
Both works release a code efficiency benchmark by crafting coding tasks, reference solutions on Python, and corresponding test cases to evaluate the performance of CodeLLMs.
Findings from EffiBench~\cite{huang2024effibench} indicate that CodeLLMs all produce code that is suboptimal in execution runtime, 
while ECCO~\cite{waghjale2024ecco} points out that existing works on improving the efficiency of CodeLLMs-generated code sacrifices their functional correctness,  
underscoring the need to consider both code efficiency and correctness in optimizing CodeLLMs.

\subsection{RLHF}

\begin{figure}[ht]
    \centering
    \includegraphics[width=0.9\textwidth]{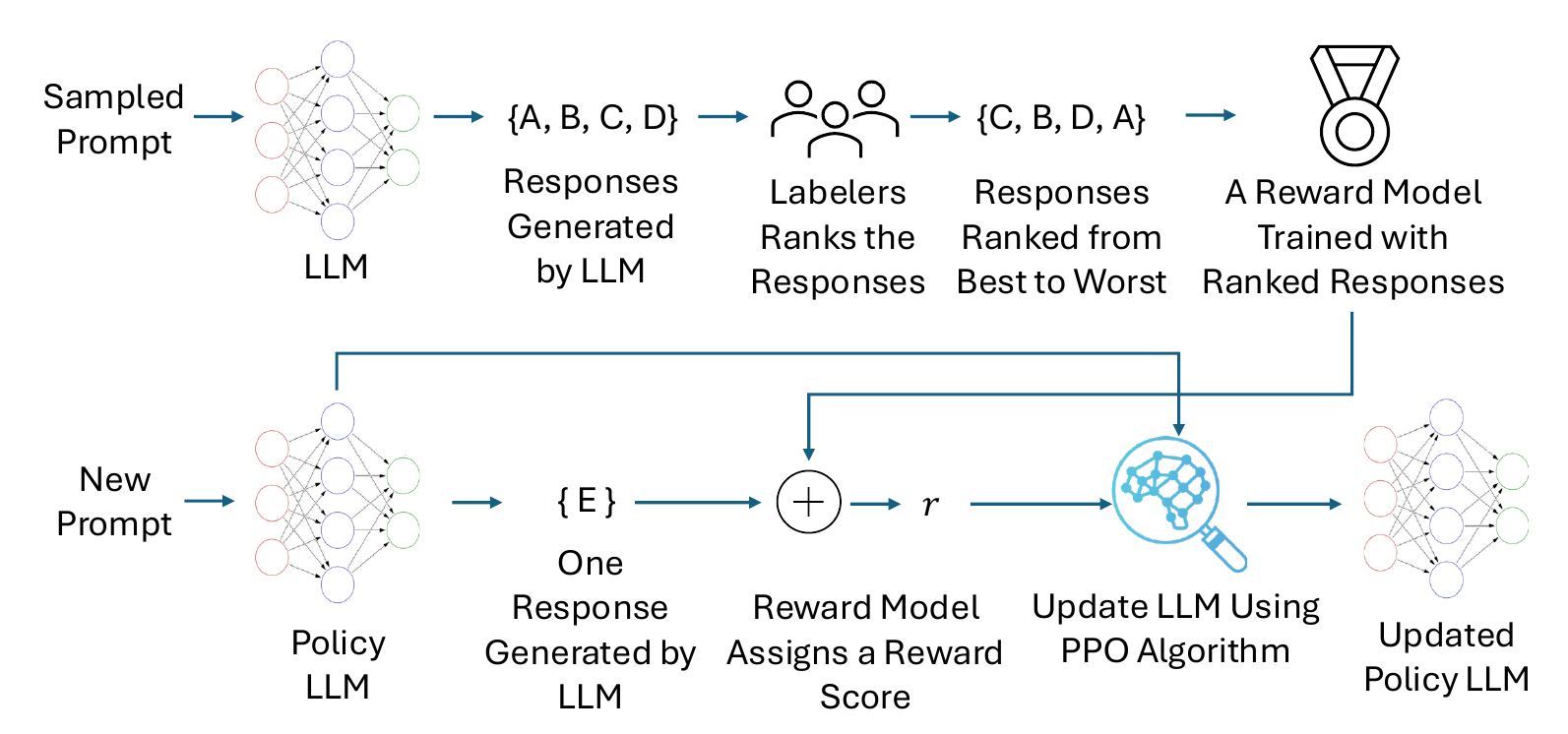}
    \caption{Overview of RLHF. Generally, the process begins with the LLM generating multiple responses for a given prompt, which are then ranked by human labelers based on their preference. These ranked responses are used to train a reward model that assigns a reward score to evaluate the human preference of newly generated responses. The LLM is then fine-tuned using the Proximal Policy Optimization (PPO) algorithm to optimize its outputs for higher reward scores, iteratively aligning LLM's outputs with human preferences.    }
    \label{fig:rlhf}
\end{figure}

Reinforcement learning from human feedback (RLHF) is a widely-used fine-tuning technique for aligning LLM outputs with human preferences~\cite{dai2023safe,ouyang2022training,wang2023aligningllm,shen2023aligningllm}. 
We introduce the overview of RLHF in Figure~\ref{fig:rlhf}.
During the pre-processing steps of RLHF, human annotators label pairwise LLM outputs as a pair of preferred and non-preferred responses, which is then used to train a rewarder that quantitatively estimates human preferences. 
During RLHF fine-tuning, the target LLM generates responses for the prompts in the training dataset. 
The rewarder receives the generated responses and assigns a scalar reward score approximating the human preferences towards the responses given the prompt.
The target LLM is then iteratively fine-tuned using a reinforcement learning algorithm, typically PPO~\cite{schulman2017proximal}, to optimize its parameters by maximizing the reward. 
% Through iterative reward maximization, the target LLM aligns its outputs more closely with human expectations. 
RLHF has proven effective in guiding LLMs to generate trustworthy, secure, and non-toxic text~\cite{dai2023safe,ouyang2022training}.
Unlike previous RLHF frameworks designed for general-purpose tasks and require manually labeled datasets to train a rewarder, ACECode specifically targets code correctness and efficiency and proposes a training-free and dual-objective rewarder grounded in code execution feedback.

\subsection{Task Definition}
We define the dual-objective code generation task as generating code solutions $\hat{C}$ that are both efficient ($G_e$) and correct ($G_c$) for a natural language instruction $I$, 
as a sequence-to-sequence problem.
This task is modeled as a sequence-to-sequence problem, where the input is a natural language instruction $I$ and the output is code snippets $\hat{C} = \left ( \hat{c}_1,..., \hat{c}_t \right )$ that satisfies the dual objectives of $G_e$ and $G_c$.

To achieve this, we optimize the model parameters $\theta$ to maximize the conditional probability of generating a correct and efficient code sequence, incorporating the dual-objective reward $R(G_e, G_c)$:

% Given the model parameters $\theta$, the optimization goal is to increase the likelihood of generating code solutions that are both efficient and correct, denoted as a conditional probability $P(C|G_e, G_c)$:

% \begin{equation}
%     \max_{} P(C|G_e, G_c, \theta) = \max  \prod_{t=1}^{T}p(C_t|C_{1:t-1}, G_e, G_c, \theta)
% \end{equation}

\begin{equation} \max_\theta \mathbb{E}_{\hat{C} \sim p_\theta(\cdot | I)} \big[ R(G_e(\hat{C}), G_c(\hat{C})) \big] \end{equation}

\noindent where $R(G_e(\hat{C}), G_c(\hat{C}))$ is a reward function explicitly quantifying code efficiency and correctness and $p_\theta(\cdot | I)$ represents the model's policy for generating code sequences. The CodeLLM is fine-tuned to adjust its policy to maximize the expected reward, aligning its outputs with the dual objectives of efficiency and correctness.

% \noindent where $C=\left ( c_1,..., c_t \right )$ refers to the ground-truth code solution, and $T$ refers to the token length of the code snippet.
% During the training process, model parameters $\theta$ are optimized to maximize the likelihood of generating code solutions that are both efficient and correct.

\section{Methodology}
\begin{figure}
    \centering
    \includegraphics[width=1\textwidth]{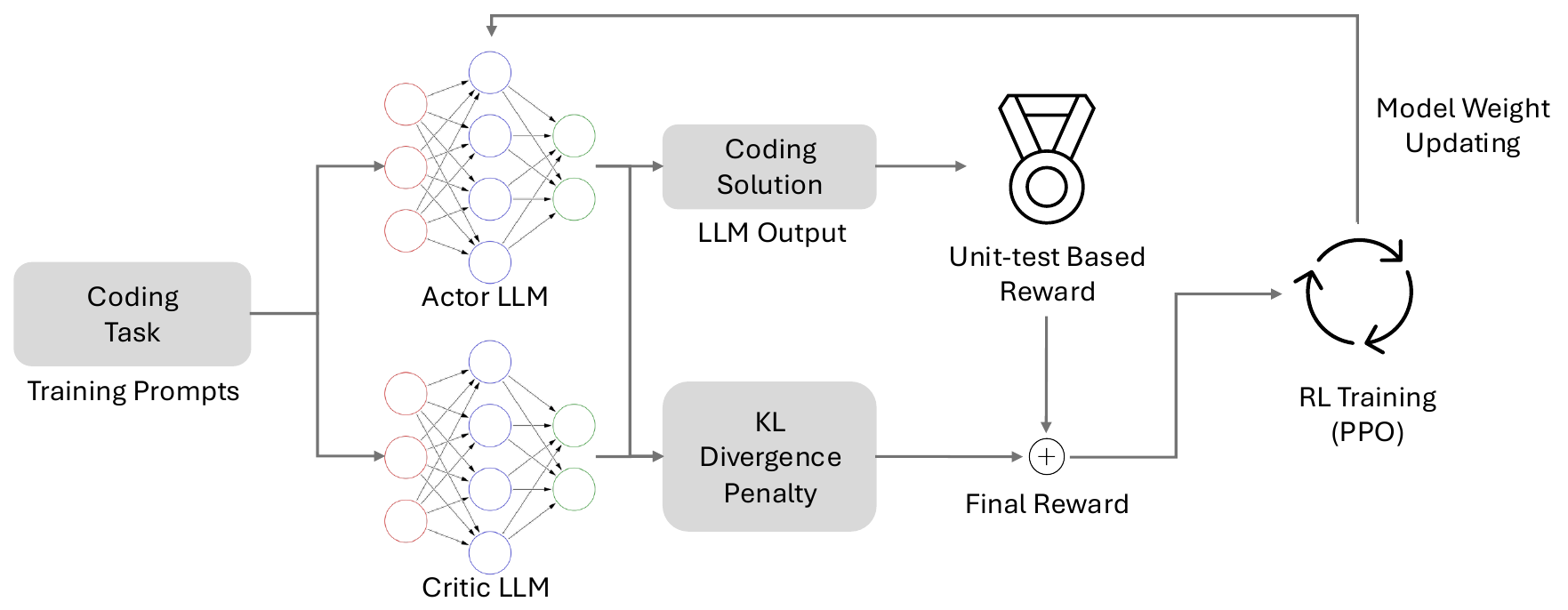}
    \caption{Overview of \tool{} with PPO training. 
    The Actor LLM, which serves as the target model for optimization, takes prompts as input and generates code snippets at each step.
    For each LLM output, the Unit-test Based Rewarder executes the generated code with all test cases, assigning a reward score to the generated code, representing both code correctness and efficiency based on execution feedback. 
    This reward score is used to guide the Actor LLM in optimizing its policy via the PPO algorithm.
    Meanwhile, the Critic LLM estimates the value function for each Actor LLM output, facilitating more stable policy optimization.
    }
    \label{fig:overview}
\end{figure}

We propose \tool{}, a novel reinforcement learning-based fine-tuning framework with dual objectives: code correctness and efficiency.
ACECode consists of three main components: (1) code generation, (2) reward calculation, and (3) policy optimization through PPO algorithm.
Figure~\ref{fig:overview} illustrates the ACECode architecture, comprising an actor LLM ($\pi_\theta$) responsible for generating code snippets, a rewarded $\mathcal{R}$ to output reward signals, and a critic LLM ($V_\pi$) to estimate value functions for stable policy updates. 
Both LLMs are optimized using the PPO algorithm~\cite{schulman2017proximal}, guided by a reward signal $\mathcal{R}$ that estimates code quality based on efficiency and correctness.
Specifically, the actor LLM serves as the target model for optimization, responsible for producing code solutions based on natural language prompts.
Meanwhile, the Critic LLM serves as a stability mechanism in the training process, estimating the value function for each code snippet generated by the Actor LLM.
The value function quantifies the expected cumulative reward of each code snippet, which stabilizes the policy updating for the actor LLM.
Finally, the tuned actor LLM ($\pi_\theta$) is extracted as the optimized target model for generating efficient and correct code solutions.

% Figure~\ref{fig:overview} illustrates the overview architecture of \tool{}.
% Specifically, \tool{} comprises an actor LLM $\pi_\theta$, responsible for code generation actions, and a critic LLM $V_\pi$, which estimates the value function. Both LLMs updates weights by Proximal Policy Optimization (PPO) algorithm~\cite{schulman2017proximal}, guided by the reward signal $R$.
% This reward signal is a scalar value that estimates code quality in terms of efficiency and correctness, steering the actor LLM to optimize its policy. Finally, the actor LLM ($\pi_\theta$) is extracted as the optimized target model.

% \tool{} has three main sub-processes: (1) code generation, (2) reward calculation, and (3) PPO training.  

\subsection{Code Generation}

The first step in ACECode’s framework is code generation. 
For each coding task $I$, the actor LLM ($\pi_\theta$) generates a code snippet $\hat{C}$. 
We apply prompts in the same format as prompts in existing benchmark of code efficiency, i.e., EffiBench~\cite{huang2024effibench}.
% We explicitly specify  requirements in prompts.
% The prompt includes open test cases for each coding task, a standard practice in code generation benchmarks~\cite{chen2021humaneval, austin2021mbpp}. 
We show the prompt template in Table~\ref{tab:prompt}.
Besides, to align the output format of the LLM with the task requirements, we add an example pair $\left( I_{example}, C_{example} \right)$ into the prompt for in-context learning. In-context learning has proven to be effective for aligning output format~\cite{brown2020fewshot}.

\begin{table}
    \centering
    \caption{Prompt templates for LLMs.}
    \label{tab:prompt}
    \begin{tabular}{p{0.8\columnwidth}}
        \toprule
        \textbf{Prompt Template For Code Generation} \\
        \midrule
        \textit{\#\#\# Instruction: Please based on the task description write Python Solution to pass the provided test cases.} \\
        \textit{You must follow the following rules:} \\
        \textit{The code should be in} \texttt{'''python\textbackslash n[Code]\textbackslash n'''} \textit{block}. \\ 
        \textit{Second, You should not add the provided test cases into the code.} \\
        \textit{Third, You do not need to write the test cases, we will provide the test cases for you.}\\
        \textit{Fourth, Only include the Python solution, without any additional explanation.}\\ 
        \textit{Fifthly, import only the necessary modules; avoid using import *.} \\
        \textit{Finally, You should make sure that the provided test cases can pass your solution.} \\
        \textit{\#\#\# Here is a example:} \\
        \textit{\{Example Instruction\}} \\
        \textit{\{Example Code Solution\}} \\
        \textit{\#\#\# Task Description:} \\   
        \textit{\{Task Description\}} \\
        
        \bottomrule 
    \end{tabular}
\end{table}

% As illustrated in Figure~\ref{fig:overview}, in the first step, the actor LLM $\pi_\theta$ generates code snippets $\hat{C}$ for each coding task $I$.
% Following the prompt from EffiBench~\cite{huang2024effibench},
% we expllicitly define the code efficiency and correctness requirements in the prompt to guide the actor LLM $\pi_\theta$ to generate efficient and correct code solutions.
% Meanwhile, 
% We also includes open test cases in the prompt, which setting is standard in code generation benchmarks~\cite{chen2021humaneval, austin2021mbpp}.
% To enable in-context learning, we add an example pair $\left( I_{example}, C_{example} \right)$ to align the output format of the LLM. The specific prompt template used is available in our GitHub repository \cran{[GitHub repo link]}. For each prompt in a training batch, the actor LLM ($\pi_\theta$) generates $N$ responses. We then use exact matching to extract the generated code snippet $\hat{C}$ from each model output. Finally, we feed a batch of $\left( C, \hat{C} \right)$ pairs into the reward calculation process for further evaluation.

The actor LLM generates $N$ responses for each input prompt.
We then extract the code snippets
$(\hat{C}_{1},...,\hat{C}_{N})$ from each response
by matching the format of the coding block that is explicitly defined in the instruction. 
Notably, 
following standard practice in RLHF~\cite{ouyang2022training, dai2023safe}, 
the temperature of LLM in the fine-tuning stage is set to 0.85 to generate diverse responses. 
$(C, \hat{C}_{1},...,\hat{C}_{N})$ is then passed to the rewarder that will assess and produce scores representing code efficiency and correctness jointly.

\subsection{Rewarder}
\label{sec:rewarder}
% The rewarder $\mathcal{R}$ is a key component of ACECode, providing a feedback signal derived from code execution to guide the actor LLM ($\pi_\theta$) toward the desired behavior—generating efficient and correct code.
The rewarder $\mathcal{R}$ serves as a proxy for code efficiency and correctness assessment, providing a scalar reward signal for each input $\hat{C}$. 
The reward signal then guides the optimization of the actor LLM ($\pi_\theta$) with dual objectives, i.e., generating efficient and correct code, through the PPO algorithm.
Different with common rewarders that are trained with human-labeled dataset~\cite{ouyang2022training, dai2023safe} and are prone to subjective bias, 
\tool{}'s
$\mathcal{R}$ applies a training-free and objective evaluator grounded by code execution, eliminating the need for manual labeling and reducing the risk of subjective bias.

Designing $\mathcal{R}$ to represent both code efficiency and correctness presents several challenges:
\begin{itemize}
    \item \textbf{C1. Requiring human-labeled dataset}: Rewarders often require a human-labeled dataset for training~\cite{ouyang2022training,dai2023safe}, which is prone to subjective bias and requires substantial labeling effort.
    \item \textbf{C2. Representing code efficiency and correctness in one rewarder}: The rewarder must encapsulate both objectives including code correctness and efficiency; otherwise, the actor LLM may prioritize one objective over the other during fine-tuning.
    \item \textbf{C3. Inaccurate runtime measurements}: 
    Some code snippets with short execution time can have high variance in the measurement of execution runtime due to potential concurrent processes 
    and can be inaccurate if measured with profiling tools~\cite{pythondoctimeit}.
\end{itemize}
The following subsections detail our approach to address these challenges.

% Designing $\mathcal{R}$ to approximate code efficiency and correctness presents three main challenges: (1) $\mathcal{R}$ often relies on models trained with human-labeled preference data~\cite{ouyang2022training, dai2023safe}, which requires substantial labeling effort and is prone to subjective bias, making it difficult to accurately capture both efficiency and correctness. (2) The reward score, represented as a scalar value, must simultaneously reflect both code efficiency and correctness. (3) Execution runtimes for short code snippets are often affected by concurrent processes, leading to high variance in runtime measurements.

\subsubsection{Objective Evaluator for Efficiency and Correctness}
% To address the first challenge, we replace the reward model, which depends on subjective human labels, with an objective evaluator based on code execution that directly measures efficiency and correctness. 
% Specifically, for a given pair $\left( C, \hat{C}\right)$, we obtain a set of ground-truth test cases $T$, where $T = \left\{ t_1, t_2, ..., t_n \right\}$.
% We execute both $\left( C, \hat{C}\right)$ on the test cases $T$, recording the compiler signals and execution runtime $ \left( {E}_{C}, {E}_{\hat{C}} \right)$ for the ground-truth and generated code solutions, respectively. 
% Follwoing EVALPERF~\cite{liu2024evaluating}, we use execution runtime as a measure of efficiency cost.
% Both compiler signals and execution runtime $ \left( {E}_{C}, {E}_{\hat{C}} \right)$ are then used to calculate the reward function.

To address challenge C1, we do not manually label code snippets with efficiency and correctness scores to train a rewarder.
Instead, we use an objective and training-free evaluator based on code execution that measures code efficiency and correctness directly as the rewarder.
Specifically, for a given pair $\left( C, \hat{C}\right)$, we obtain a set of ground-truth test cases $T = \left\{ t_1, t_2, ..., t_n \right\}$ and execute both $C$ and $\hat{C}$ on $T$. 
We record the compiler signals, which indicate whether the code encounters a compilation error, and execution runtimes $\left( E_{C}, E_{\hat{C}} \right)$ for the ground-truth and generated solutions, respectively. 
We then apply a customized reward function $\mathcal{R}$ to calculate the reward score that reflects both code efficiency and correctness based on the compiler signals and runtime values.

% These compiler signals and runtime values are then used to calculate the reward function.

\subsubsection{Reward Function}
\label{sec:reward_function}
To address challenge C2, we design $\mathcal{R}$ as a step function that evaluates the generated code's correctness and efficiency based on compiler outcomes and runtime performance:

\begin{equation}
    \mathcal{R}(\hat{C}, C) = 
    \begin{cases} 
    0.5 + 0.5 \cdot \min\left(\left(\frac{{E}_{C}^{t}}{{E}_{\hat{C}}^{t}}\right)^k, 1\right), & \text{if } \hat{C} \text{ passes all unit tests} \\
    -0.3, & \text{if } \hat{C} \text{ fails any unit test} \\
    -0.5, & \text{if } \hat{C} \text{ encounters a compile error} 
    \end{cases}
    \end{equation}

\noindent where
%  ${E}_{C}^{t}$ and ${E}_{\hat{C}}^{t}$ represents the execution runtime for ground-truth and generated code solution, respectively. 
% $k$ is a penalty factor to control the sensitivity of efficiency loss of $\hat{C}$ relative to $C$.
$k$ is a penalty factor controlling the sensitivity to efficiency gain. 
% A penalty hyperparameter is necessary to achieve a balance between the gains in efficiency and correctness. 
% The improvements in efficiency may not be linear, for example when the generated solutions differ in their asymptotic complexity, and the exponential rate of improvement would result in a reward that is dominated by the efficiency improvements.  
% \hj{I tried to write something late at night, but maybe it doens't make sense. Check again later}
% \textcolor{red}{
% We represent efficiency gain with time complexity
% We introduce $k$ to amplify the improvement rates from runtime ratio to time complexity. 
% We argue 
% When the runtime ratio $\frac{{E}_{C}^{t}}{{E}_{\hat{C}}^{t}}$ grows linearly as the generated code becomes faster relative to the ground-truth solution, improvements in time complexity (e.g., from $ ( O(n^2)) $ to $( O(n \log n) )$ produce exponential gains as input size increases, $k$ serves to align these differing scales.
% A higher $k$ value can amplify the reward for substantial efficiency improvements, making it more comparable to the exponential impact of changes in time complexity.
% }
% \cran{btw, this is the initial concern for me to set a penalty factor but I forgot to write it...Do you think this makes sense to you? And if yes, I think we may consider it as a contribution and need one more ablation study on that.} Actually, I am not so sure if this makes sense, so the explanation may need improvements to be clearer. 

This reward structure encourages $\pi_\theta$ to generate code that is both correct and efficient. 
Following existing approaches~\cite{dai2023safe, ouyang2022training}, we normalize the reward values from $-1$ to $1$ to stabilize the training process.
Firstly, if $\hat{C}$ fails to compile or pass any unit test, it receives a negative reward.
Code that fails to compile receives a penalty of $-0.5$, set as half the worst-case value to avoid making the model overly cautious and under-optimizing. A smaller penalty of $-0.3$ is assigned for code that fails any unit test, as compilation correctness is a prerequisite for functional correctness.
Conversely, if $\hat{C}$ passes all tests, it receives a positive reward comprising a fixed score $0.5$ representing code correctness and an adaptive score $0.5 \cdot \min\left(\left(\frac{{E}_{C}^{t}}{{E}_{\hat{C}}^{t}}\right)^k, 1\right)$ representing code efficiency.
To encourage balanced optimization of both objectives, the fixed score for correctness is set to $0.5$ and the adaptive score for efficiency is capped at a maximum of $0.5$.
Concretely, the code efficiency of $\hat{C}$ is quantified as the ratio of its runtime ${{E}_{\hat{C}}^{t}}$ to that of the ground-truth ${E}_{C}^{t}$.

% fixed score of $0.5$ for correctness and an adaptive score for efficiency. The adaptive score for efficiency is capped at a maximum of $0.5$ to ensure balanced optimization of both objectives.

% When $\hat{C}$ passes all tests, the reward score includes a fixed score of $0.5$ for correctness rewarding and an adaptive score for efficiency rewarding.

% Specifically, if $\hat{C}$ fails to compile or pass any unit test, it receives a negative reward, with a larger penalty assigned for compile errors, as compile correctness is a prerequisite for execution.
% When $\hat{C}$ passes all tests, the reward score includes a fixed score of $0.5$ for correctness rewarding and an adaptive score for efficiency rewarding.

\subsubsection{Precise Runtime Measurement}
Existing benchmark EffiBench~\cite{huang2024effibench} applies profiling tool \texttt{memory-profiler}\footnote{\url{https://pypi.org/project/memory-profiler/}} to execute code once and extract the runtime.
However, as mentioned by official Python documentation\footnote{\url{https://docs.python.org/3.9/library/profile.html}}, profiling tools would introduce additional overhead to the runtime measurement of Python code and are not designed for benchmarking purposes. Meanwhile, we observe that the runtime of each code may be affected by concurrent processes, resulting in inconsistent runtime for each execution. 

Therefore, to address challenge C3,
following recommendations from the official Python documentation\footnote{\url{https://docs.python.org/3.9/library/profile.html}},
we use Python's \texttt{timeit} module~\cite{pythondoctimeit}. 
This allows reasonable assessments of execution runtime with less overhead. 
Additionally, to mitigate the effect of randomness in the runtime between different executions, we implement a repeated execution strategy.
We calculate the runtime value 
by executing each generated code snippet $n$ times and computing the average, $\widetilde{E}_{\hat{C}^t}$. 
% This approach reduces runtime variability. 
We define two parameters for repeating execution:
\begin{itemize}
    \item $n_{min}$: the minimum number of executions needed to stabilize runtime measurements.
    \item $t_{max}$: the maximum allowable accumulated runtime to avoid an infinite loop.
\end{itemize}
The total number of executions ($n$) is then calculated as:

\begin{equation}
    \label{eq:repetition}
    n = \max\left( n_{min}, \frac{t_{max}}{n_{t_{max}}} \right)
\end{equation}

\noindent where $n_{t_{max}}$ is the number of completed executions when the accumulated runtime reaches $t_{max}$.

We also consider the overhead of importing libraries.
We move all import statements to the \texttt{setup} parameter of \texttt{timeit}, ensuring they are executed only once and excluded from measurement cycles.

Finally, for all input code $(\hat{C}_{1},...,\hat{C}_{N})$, the rewarder assigns reward values, which are then used to guide the PPO training.

\subsection{PPO Training Framework}
% Inspired by ChatGPT~\cite{ouyang2022training}, we propose a reinforcement learning framework to fine-tune CodeLLMs specifically for generating efficient and correct code.
We use the PPO algorithm~\cite{schulman2017proximal} to fine-tune the actor LLM $\pi_\theta$ with the reward signal $R$.
% We employ the Proximal Policy Optimization algorithm~\cite{schulman2017proximal} to fine-tune the actor model $\pi_\theta$, using a reward signal $\mathcal{R}$ that reflects code quality in terms of both efficiency and correctness.
This code generation process is formulated as a sequential discrete Markov Decision Process, where each step in the generation process is defined by tuple $\left ( State, Action, Policy, Reward \right )$ and solved by PPO. 
At each time step $t$, representing the token length of the generated code so far, the components of the MDP process are defined as follows:

\begin{itemize}
    \item \textbf{State} $S_t$: The state at time step $t$ consists of the input prompt $I$ and the partially generated code snippet $\hat{C}_{1
    } = \left( \hat{c}1, \ldots, \hat{c}{t-1} \right)$. This concatenation forms the input context up to token $t$.    
    \item \textbf{Action} $A_t$: The action at each time step $t$ is defined as generating the next token $\hat{c}_t$ in the code sequence.
    \item \textbf{Policy} $\pi_\theta$: The policy is represented by actor LLM $\pi_\theta$, which performs the action $A_t$, i.e., generates the next token $\hat{c}_{t}$ given the state $S_t$.
    \item \textbf{Rewarder} $\mathcal{R}_t$: The rewarder $\mathcal{R}_t$ is applied at the end of each generation episode, i.e., when the complete code snippet $\hat{C}$ is produced. 
    The rewarder produces a scalar reward value that reflects both code correctness and efficiency, as detailed in Section~\ref{sec:rewarder}.

    \item \textbf{Advantage} $A_t$: The advantage $A_t$ quantifies the value of the action $A_t$ relative to the critic LLM's estimated value function $V_\pi$, given the current state $S_t$. It is computed using the critic model $V_\pi$ and the reward signal $\mathcal{R}_t$. Specifically, for each time step $t$, we calculate $A_t$ as follows:
    
    \begin{equation}
        A_t = \mathcal{R}_t + \gamma V(s_{t+1}) - V(s_t)
    \end{equation}

    \noindent where $V(s_t)$ is the value of the current state as estimated by the critic LLM and $\gamma$ is a discount factor.

\end{itemize}

The objective of training is to maximize the expected reward by optimizing the policy $\pi_\theta$. This objective can be formalized as:

\begin{equation}
    \max_{\theta} \; \mathbb{E}_{I \sim \mathcal{I}, \; \hat{c} \sim \pi_{\theta}(\cdot \mid I)} \left[ \mathcal{R}(\hat{c}, I, c) \right]
\end{equation}

\noindent where $\mathcal{I}$ is the dataset of input prompts, and $c$ is the ground-truth code for each prompt.
Specifically, the PPO loss function $L_{\text{PPO}}(\theta)$ is defined as:

\begin{equation}
% L_{\text{PPO}}(\theta) = \mathbb{E} \left[ \min \left( r_t(\theta) A_t, \text{clip}(r_t(\theta), 1 - \epsilon, 1 + \epsilon) A_t \right) \right]
L_{\text{PPO}}(\theta) = \mathbb{E} \left[ \min \left( r_t(\theta) A_t, \text{clip}\left(r_t(\theta), 1 - \epsilon, 1 + \epsilon\right) A_t \right) \right]
\end{equation}

\noindent where $r_t(\theta) = \frac{\pi_\theta(a_t \mid s_t)}{\pi_{\theta_{\text{old}}}(a_t \mid s_t)}$ is the probability ratio between the current and previous policies for action $a_t$. $\text{clip}(\cdot, 1 - \epsilon, 1 + \epsilon)$ ensures the ratio is constrained within a small range to prevent large updates, with $\epsilon$ being a hyperparameter. Concretely, the $\text{clip function}$ is defined as follows:

\begin{equation}
    \text{clip}(r_t(\theta), 1-\epsilon, 1+\epsilon) = 
    \begin{cases} 
    1-\epsilon, & \text{if } r_t(\theta) < 1-\epsilon, \\
    r_t(\theta), & \text{if } 1-\epsilon \leq r_t(\theta) \leq 1+\epsilon, \\
    1+\epsilon, & \text{if } r_t(\theta) > 1+\epsilon.
    \end{cases}
    \end{equation}

In parallel, the critic LLM $V_\pi$ is updated by minimizing the following value loss:

\begin{equation}
    L_{\text{value}}(\pi) = \mathbb{E} \left[ \left( V_{\pi}(s_t) - \left( R_t + \gamma V_{\pi}(s_{t+1}) \right) \right)^2 \right]
\end{equation}

\noindent where $V_\pi(s_t)$ is the value estimate for the current state; $\mathcal{R}t + \gamma V\pi(s_{t+1})$ serves as the target value. 

To prevent overfitting during policy updates, we incorporate a per-token KL divergence penalty into the PPO objective, which is demonstrated to be effective in prior work~\cite{ouyang2022training}. This adjustment controls how closely the updated policy adheres to the previous policy, $\pi_{\theta_{\text{old}}}$, and the combined loss for PPO is expressed as: 

\begin{equation}
    L_{\text{total}} = L_{\text{PPO}}(\theta) + \lambda L_{\text{value}}(\phi) - \beta \text{KL}(\pi_{\theta} \, \| \, \pi_{\theta_{\text{old}}})
    \end{equation}

\noindent where $\lambda$ is a weighting coefficient for the value loss and $\beta$ is the coefficient for the KL penalty, which mitigates excessive divergence between the current and old policies.
% \begin{equation}
% \max_{\theta} \; \mathbb{E}{I \sim \mathcal{I}, \; \hat{c} \sim \pi{\theta}(\cdot \mid I)} \left[ \mathcal{R}(\hat{c}, I, c) - \beta \; \text{KL}\left(\pi_{\theta_{\text{old}}}(\cdot \mid I) \, \| \, \pi_{\theta}(\cdot \mid I) \right) \right]
% \end{equation}

% \noindent Where $\beta$ is coefficient to control the KL penalty strength, and $\pi_{\theta_{\text{old}}}$ is the policy model before the current update step.
% The PPO algorithm iteratively updates the policy $\pi_\theta$ across training batches until convergence. The final actor policy $\pi_\theta$ is then extracted as the optimized target model for generating efficient and correct code solutions.

\section{Experimental Setup}
\label{sec:experiment_setting}
\subsection{Dataset}
\label{sec:dataset}
To evaluate \tool{}, we use an extended version of the EffiBench~\cite{huang2024effibench} dataset, which is designed for assessing code efficiency in CodeLLMs.
EffiBench consists of 1,000 efficiency-focused coding tasks extracted from LeetCode~\cite{leetcode}, each paired with a single ground-truth solution.

To compute the reward for a CodeLLM-generated code solution, \tool{} determines the level of code efficiency by comparing its execution runtime against the ground-truth reference solution, which ideally serves as an upper bound for runtime efficiency.
% Therefore, \tool{} requires the reference solution to be as efficient as possible.
While EffiBench extracts one most-upvoted solution from LeetCode’s discussion forum as a reference solution for each coding task, we observe that these solutions are not always the most runtime-efficient, as some prioritize readability or memory usage over runtime efficiency.
To address this, we extend EffiBench by gathering at most 15 most-upvoted solutions for each coding task, executing them, and then picking the solution with the shortest runtime as a new reference solution.
This approach establishes an accurate benchmark for the upper bound efficiency of code solutions for each coding task, allowing for a more precise calculation of code efficiency reward and fair evaluation.
Note that coding tasks that are restricted to premium users or contain faulty test cases are excluded from our dataset. The resulting extended dataset, which we refer to as EffiBench+, comprises 797 coding tasks and 8,520 code solutions.
% Each coding task in EffiBench+ is paired with 10 to 15 human-written and ground-truth solutions. 
We randomly split EffiBench+ into training and test sets with an 8:2 ratio. This test set aligns in size with HumanEval~\cite{chen2021humaneval}, one of the most widely used benchmarks for code generation.

\subsection{Selecting CodeLLMs}
\label{sec: llm_selection}

We fine-tune four state-of-the-art (SOTA) open-source CodeLLMs selected from the top-performing models on the Evalplus~\cite{evalplus} leaderboard. Specifically, we choose SOTA models including Magicoder~\cite{wei2023magicoder}, DeepSeekCoder~\cite{zhu2024deepseek}, CodeQwen1.5~\cite{bai2023qwen}, and WaveCoder~\cite{yu2023wavecoder}. Specifically, we select the model version with parameters less than 10B due to the computational constraints of our training environment.
We give the details of each selected CodeLLM in Table~\ref{tab:code_models}.
Note that as \tool{} requires updating the internal states of CodeLLMs via reinforcement learning, we are unable to evaluate commercial closed-source models.

\begin{table}[ht]
    \centering
    \caption{Summary of Selected CodeLLMs}
    \resizebox{\textwidth}{!}{%

    \begin{tabular}{ccc}
    \toprule
    \textbf{Model Name} & \textbf{Number of Parameters} & \textbf{Institution} \\ 
    \midrule
    DeepseekCoder-6.7B-instruct      & 6.7B                   & DeepSeek AI           \\ 
    Magicoder-S-DS-6.7B           & 6.7B                   & Intelligent Software Engineering-UIUC \\ 
    CodeQwen1.5-7B-Chat         & 7B                           & Alibaba Cloud         \\ 
    WaveCoder-Ultra-6.7B           & 6.7B             & Microsoft             \\ 
    \bottomrule
    \end{tabular}
    }
    \label{tab:code_models}
    \end{table}

Magicoder~\cite{wei2023magicoder} is a family of CodeLLMs designed for code generation tasks.
These models are trained using OSS-Instruct, a novel dataset that leverages open-source code snippets to generate diverse and realistic instruction data. 
This methodology aims to mitigate inherent biases present in LLM-synthesized data by incorporating a wealth of open-source references, resulting in more diverse, realistic, and controllable outputs.

DeepSeekCoder~\cite{zhu2024deepseek} is an open-source series of CodeLLMs that are designed to enhance code generation and understanding tasks. The base model of DeepSeekCoder is trained from scratch on a massive dataset of 2 trillion tokens, comprising 87\% code and 13\% natural language in both English and Chinese.
They utilize a 16K context window and a fill-in-the-blank task to support project-level code completion and infilling tasks.

CodeQwen1.5~\cite{bai2023qwen} is a code-specific variant of the Qwen1.5 LLM, which is pre-trained on approximately 3 trillion tokens of code-related data, encompassing 92 programming languages. The model supports a context length of up to 64,000 tokens.

WaveCoder~\cite{yu2023wavecoder} is a series of CodeLLMs fine-tuned on an instruction dataset called CodeSeaXDataset, which comprises 19,915 instruction instances across four tasks: code generation, code summarization, code translation, and code repair. The fine-tuning process employs a generator-discriminator framework to ensure high-quality and diverse instruction data.

\subsection{Implementation Details}
We implement \tool{} based on the HuggingFace Transformers platform~\cite{huggingface} and OpenRLHF framework~\cite{hu2024openrlhf}. 
% Prior to the training phase, we execute all 10–15 human-written ground-truth solutions for each coding task and select the solution with the shortest execution time as the ground-truth solution for reward calculation.

To ensure consistent reward calculation, we execute each ground-truth and CodeLLM-generated solution pair consecutively on the same machine, reducing potential variance introduced by the execution environment.
Meanwhile, we observe that more than 95\% ground-truth solutions in EffiBench require less than 1 seconds to execute in our local environment, so we set the maximum allowable accumulated execution time ($t_{max}$) in equation~\ref{eq:repetition} to 3 seconds. We also set the execution count to 10 to ensure stable evaluation results. 
Furthermore, any code execution exceeding 10-seconds timeout would be terminated to avoid infinite loops.
% We observes that almost ground-truth solutions we collected for EffiBench requries less than 3 seconds to execute, therefore we set the maximal accumulated execution time $t_{max}$ as 3 seconds.
To facilitate reproducibility, we set the generation parameters with a temperature of 0 and top-p of 1 during the inference stage, ensuring deterministic outputs from the CodeLLMs for identical prompts.

During the RLHF training phase, 
we separate the parameters of actor and critic networks. 
As recommended in OpenRLHF~\cite{hu2024openrlhf}, we set the training batch size as 64, rollout batch size as 512, maximum generate length as 1024, learning rate for actor model as $5e^{-7}$, learning rate for critic model as $9e^{-6}$, and KL penalty coefficient as $0.01$.
Following standard practice in RLHF~\cite{xu2024dpo,hu2024openrlhf}, we configure the sampling strategy of LLM by setting the temperature and top-p as 1, and we sample 5 responses for each prompt in training. 
To accelerate training, we apply FlashAttention~\cite{dao2022flashattention} and DeepSpeed~\cite{rasley2020deepspeed}. We randomly sample 5\% of the training data as a validation set and select the checkpoint with the best validation performance for final evaluation.

\subsection{Baselines}
We consider three baseline settings for evaluating the effectiveness of \tool{}: 
\begin{itemize}
    \item \textbf{Original CodeLLMs}: We evaluate the correctness and efficiency of the code from CodeLLMs fine-tuned with \tool{} against the original CodeLLMs without fine-tuning.
    % \item \textbf{CodeLLMs with Few-shot Prompting}: We construct one baseline by prompting CodeLLMs with few-shot prompting~\cite{brown2020fewshot}. Concretely, we follow PIE~\cite{shypula2023learning} to prompt CodeLLMs with a pair of efficient and inefficient code examples that are extracted from training dataset for each coding task. The detailed prompt is shown in Table \ref{tab:fewshot}.
    % \item chain-of-thought prompting (refered as COT)~\cite{wei2022chainofthoughtcot}. \textcolor{red}{check it later.}
    \item \textbf{CodeLLMs with Instruction Tuning}: We perform instruction tuning on the original CodeLLMs using the same training split from EffiBench as \tool{} and compare their performance with CodeLLMs fine-tuned by \tool{}. 
    % We use the original dataset format of EffiBench, in which each coding task is paired with a single ground-truth solution. 
    The instruction is constructed with the task description, a one-shot example of task description and code, and the ground-truth solution. The template of the task description is the same as \tool{}, which is illustrated in Table~\ref{tab:prompt}. 
    We fine-tune CodeLLMs using the instructions with the next token prediction objective.
    \item \textbf{PIE~\cite{shypula2023learning}}: 
    We consider PIE as a strong baseline for code efficiency optimization.
    PIE applies instruction tuning with performance-conditioned generation to optimize the efficiency of CodeLLM-generated code. Performance-conditioned generation involves tagging code solutions with performance scores during training (e.g., the top 10\% most efficient solutions are labeled as "10/10," the next 10\% as "9/10," and so on). PIE's dataset consists of pairs of efficient and inefficient code snippets for each task, written in C++. 
    During the training stage, the ground-truth performance scores are provided for each input code to quantify the extent of code efficiency. During the inference stage, PIE prompts the LLM to generate a code solution with a 10/10 performance score.

    As the released model checkpoints of PIE are optimized for C++, and our training dataset is focused on Python, we extend PIE approach to Python for a fair comparison. Specifically, we follow PIE's instruction-tuning methodology while leveraging our extended version of EffiBench (see Section~\ref{sec:dataset}). Our extension provides multiple ground-truth code solutions per coding task, which are executed once before training to assign performance tags. 

\end{itemize}

% \begin{table}
%     \centering
%     \caption{Prompt templates for Few-shot Prompting.}
%     \label{tab:prompt}
%     \begin{tabular}{p{0.8\columnwidth}}
%         \toprule
%         \textbf{Prompt Template For Code Generation} \\
%         \midrule
%         \textit{\#\#\# Instruction: Please based on the task description write Python Solution to pass the provided test cases.} \\
%         \textit{You must follow the following rules:} \\
%         \textit{The code should be in} \texttt{'''python\textbackslash n[Code]\textbackslash n'''} \textit{block}. \\ 
%         \textit{Second, You should not add the provided test cases into the code.} \\
%         \textit{Third, You are not need to write the test cases, we will provide the test cases for you.}\\
%         \textit{Fourth, Only include the Python solution, without any additional explanation.}\\ 
%         \textit{Fifthly, import only the necessary modules; avoid using import *.} \\
%         \textit{Finally, You should make sure that the provided test cases can pass your solution.} \\
%         \textit{\#\#\# Here is a example:} \\
%         \textit{\{Example Instruction\}} \\
%         \textit{\{Example Code Solution\}} \\
%         \textit{\#\#\# Task Description:} \\   
%         \textit{\{Task Description\}} \\
        
%         \bottomrule 
%     \end{tabular}
% \end{table}

\subsection{Evaluation Metrics}
\label{sec: eval_metrics}
\subsubsection{Code Correctness}
we use the pass@1 metric to evaluate the code correctness of CodeLLMs. The pass@1 metric measures the proportion of generated code solutions that pass all test cases on the first attempt. Pass@1 ranges from 0 to 1, with higher values indicating better code correctness:

\begin{equation} 
    \text{pass@1} = \frac{\text{Number of passed solutions}}{\text{Total number of coding tasks}} 
\end{equation}

\subsubsection{Code Efficiency}
Existing benchmarks for evaluating code efficiency, such as EffiBench~\cite{huang2024effibench}, use metrics like Average Execution Time (AET) and Normalized Average Execution Time (NAET). AET calculates the arithmetic mean execution time of generated code solutions, while NAET measures the relative arithmetic mean execution time of generated solutions compared to the ground-truth solution.
However, we empirically find that AET and NAET are disproportionately influenced by outliers because they compute the arithmetic mean.
% However, we argue that both AET and NAET may be biased due to their reliance on the arithmetic mean, which can be disproportionately influenced by outliers in execution times across coding tasks. 
For instance, if a generated code solution for a particular task has an exceptionally longer execution time than other tasks, it may dominate the AET, skewing the metric. Similarly, if one solution’s inefficiency relative to the ground-truth solution is significantly higher than other tasks, NAET would also become biased toward this outlier.

To mitigate this bias, we propose a new metric called Efficiency Code Counting (ECC), inspired by differential testing~\cite{differentialtestingsurvey}. ECC calculates the percentage of code solutions generated by CodeLLMs fine-tuned with ACECode that exhibit lower execution times than those generated by the original CodeLLM for the same tasks and test cases. Formally, ECC is defined as:

\begin{equation}
    \label{eq:ecc}
    \text{ECC} = \frac{1}{N} \sum_{i=1}^{N} \mathbf{1} \left(t^{\text{originLLM}}_{i} > t_i\right)
\end{equation}

\noindent where $N$ is the number of code generated by CodeLLM, $t_i$ is $i$-th code task, and $t^{\text{originLLM}}_{i}$ is the accumulated execution time of the original CodeLLM generated solutions. 
The ECC metric ranges from 0 to 1 and is not affected by extreme outliers. 
An ECC value above 0.5 indicates that more than half of the solutions generated by the fine-tuned CodeLLM are more efficient than those of the original model.

One limitation of ECC is that it only counts the number of more efficient solutions generated by the fine-tuned CodeLLM without accounting for the magnitude of the efficiency improvement. To address this, we introduce one additional metric: Geometric Execution Time (GET), inspired by recent works on accelerating software testing~\cite{wang2021faster}. Unlike AET and NAET which apply arithmetic means of execution time, GET calculates the geometric mean of execution time. The geometric mean allows us to capture the average improvement in efficiency without undue influence from outlier values. 
Specifically, we define GET as follows:

\begin{equation}
    \label{eq:get}
    GET = \sqrt[N]{\prod_{i=1}^{N} \frac{t_i}{n_i}}
\end{equation}

\noindent where $N$ is the number of code tasks, $t_i$ is the accumulated execution time of the code solution for $i$-th task, and $n_i$ is the number of code executions through adaptive repeated execution strategy. GET calculates the geometric mean of execution time for all code tasks. A lower GET value of the fine-tuned CodeLLM relative to the original CodeLLM indicates better runtime efficiency of the generated code.

To ensure fair and consistent comparison with EffiBench, we evaluate code efficiency only on tasks for which both the fine-tuned and original CodeLLM generate correct solutions. In this paper, we consider both ECC and GET as the primary metrics for assessing code efficiency improvements.

\section{Experimental Results}
This section describes the evaluation results of CodeLLM fine-tuned by \tool{} compared to the original CodeLLM.
Specifically, our evaluation focuses on the following two research questions (RQ):

\begin{itemize}
    \item \textbf{RQ1:} How do the CodeLLMs fine-tuned by \tool{} perform in terms of code correctness and efficiency?
    \item \textbf{RQ2:} How do the parameters of \tool{} affect the performance of the fine-tuned CodeLLMs?
\end{itemize}

\subsection{RQ1: Code Correctness and Efficiency of CodeLLMs}
In this RQ, we fine-tune CodeLLMs by \tool{} and evaluate the correctness and efficiency of the generated code. 
We compare the performance of these models with the original CodeLLMs, instruction-tuned CodeLLMs, and CodeLLMs fine-tuned using PIE~\cite{shypula2023learning}.

% We compare the performance of CodeLLMs fine-tuned by \tool{} with the original CodeLLMs, instruction-tuned CodeLLMs, and CodeLLMS fine-tuned by PIE~\cite{shypula2023learning}, respectively. The retional of separate evaluation is due to the coding efficiency being evaluated on the common set of successfully generated codes, which are not the same for different fine-tuning methods. 

\begin{table}[h]
    \caption{Correctness and Efficiency of Code Generated by Original CodeLLMs and CodeLLMs Fine-tuned by \tool{}.} 
    \begin{tabular}{llll}
    \hline
                & ECC (\%) & GET \textdownarrow & Pass@1 (\%) \textuparrow \\ \hline
        CodeQwen1.5-7B-Chat & N/A    & $1.121 \times 10^{-3}$  & 41.94 \\
        \: + \tool{}   & 72.3\%    & $1.065 \times 10^{-3}$ (\textcolor{red}{\textdownarrow 4.98\%}) & 46.77 (\textcolor{red}{\textuparrow 11.52\%})   \\ \hline
        DeepseekCoder-6.7B-instruct   & N/A    &  $8.551 \times 10^{-4}$  & 44.44   \\ 
        \: + \tool{}   & 61.2\%    & $8.505 \times 10^{-4}$ (\textcolor{red}{\textdownarrow 0.52\%}) & 48.41 (\textcolor{red}{\textuparrow 8.93\%})   \\ \hline
        WaveCoder-Ultra-6.7B & N/A    & $1.749 \times 10^{-3}$    & 42.86 \\
        \: + \tool{}   & 65.4\%    & $1.559 \times 10^{-3}$ (\textcolor{red}{\textdownarrow 10.86\%})  & 43.65 (\textcolor{red}{\textuparrow 1.84\%})   \\ \hline
        Magicoder-S-DS-6.7B & N/A    & $1.152 \times 10^{-3}$ & 49.21 \\
        \: + \tool{}   & 67.9\%     & $1.063 \times 10^{-3}$ (\textcolor{red}{\textdownarrow 7.69\%})  & 56.35 (\textcolor{red}{\textuparrow 14.51\%})   \\ \hline
    
    \end{tabular}
    \label{tab:evaluation}
    \begin{flushleft}
        \small
        \textit{As defined in Section~\ref{sec: eval_metrics}, ECC measures the percentage of code generated by tuned CodeLLMs that has less execution time than that of original CodeLLMs. GET measures the geometric average execution time of the generated code. Pass@1 measures the percentage of generated code that passes the test cases at the first attempt.}
        \end{flushleft}
    \end{table}

\subsubsection{Comparing With Original CodeLLMs}
We show the evaluation results of CodeLLMs fine-tuned by \tool{} against the original CodeLLMs in Table~\ref{tab:evaluation}.
We observe that all CodeLLMs fine-tuned by \tool{} outperform the original CodeLLMs in terms of both code correctness and efficiency.
Specifically, the ECC score of all CodeLLMs fine-tuned by \tool{} is higher than 60\%, indicating that at least 60\% of the code generated by all CodeLLMs fine-tuned by \tool{} has executes in a shorter time than the original CodeLLMs. 
Similarly, GET for all CodeLLMs fine-tuned by \tool{} decreased from 0.52\% to 10.86\% compared to the original CodeLLMs.
Both GET and ECC scores illustrate the superior efficiency of the code generated by CodeLLMs fine-tuned by \tool{}.
Moreover, we can also observe that the pass@1 score of all CodeLLMs fine-tuned by \tool{} is higher than the original CodeLLMs by from 1.84\% to 14.51\%, showing the improvement in code correctness of the generated code.

Besides, we observe that the less efficient the original CodeLLMs, the more significant the efficiency improvements achieved by \tool{}. 
% Since GET measures the geometric average execution time of generated code, a higher GET score for the original CodeLLM indicates less code efficiency. 
For all CodeLLMs in this study, the higher the original CodeLLM's GET score, the more substantial the efficiency gains achieved by \tool{}.
We attribute this trend to an upper bound on achievable efficiency improvements. 
CodeLLMs that are less efficient offer more room for targeted fine-tuning improvements.
% \hj{The text seems quite repetitive in these few paragraphs, but if you want to include them to have more pages for now, I think it's ok}

\begin{table}[]
    \caption{Correctness and Efficiency of Code Generated by CodeLLMs with Instruction tuning and CodeLLMs Fine-tuned by \tool{}.}

    \begin{tabular}{@{}llllll@{}}
    \toprule
    Model Name                                   & Fine-Tune & ECC (\%)   & GET  & Pass@1 (\%) \\ \midrule
    \multirow{2}{*}{CodeQwen1.5-7B-Chat}         &   \makecell[l]{Instruction\\Tuning}        & 47.1   & $3.03 \times 10^{-4}$ & 30.95       \\ \cmidrule(l){2-5} 
                                 &  \tool{}         & 72.3 & $2.72 \times 10^{-4}$ (\textcolor{red}{\textdownarrow 10.23\%}) & 46.77 (\textcolor{red}{\textuparrow 51.15\%})      \\ \cmidrule(l){1-5} 
    \multirow{2}{*}{DeepseekCoder-6.7B-instruct} & \makecell[l]{Instruction\\Tuning}          & 56.7   & $1.21\times 10^{-3}$ & 42.06       \\ \cmidrule(l){2-5} 
                                                 &  \tool{}          & 61.2 & $1.09\times 10^{-3}$ (\textcolor{red}{\textdownarrow 9.92\%}) & 48.41 (\textcolor{red}{\textuparrow 15.10\%})      \\ \cmidrule(l){1-5} 
    \multirow{2}{*}{WaveCoder-Ultra-6.7B}        & \makecell[l]{Instruction\\Tuning}          & 48.1   & $9.82\times 10^{-4}$ & 41.26       \\ \cmidrule(l){2-5} 
                                                 &  \tool{}          & 65.4 & $8.66\times 10^{-4}$ (\textcolor{red}{\textdownarrow 11.81\%}) & 43.65 (\textcolor{red}{\textuparrow 5.79\%})      \\ \cmidrule(l){1-5} 
    \multirow{2}{*}{Magicoder-S-DS-6.7B}         & \makecell[l]{Instruction\\Tuning}          & 60.8   & $9.19\times 10^{-4}$ & 50.00       \\ \cmidrule(l){2-5} 
                                                 &  \tool{}          & 69.7 & $7.06\times 10^{-4}$ (\textcolor{red}{\textdownarrow 23.18\%}) & 56.35 (\textcolor{red}{\textuparrow 12.70\%})      \\ \bottomrule
\end{tabular}
\label{tab:evaluation_instruct}
\begin{flushleft}
    \small
    \textit{As defined in Section~\ref{sec: eval_metrics}, ECC measures the percentage of code generated by tuned CodeLLMs that has less execution time than that of original CodeLLMs. GET measures the geometric average execution time of the generated code. Pass@1 measures the percentage of generated code that passes the test cases at the first attempt.}
    \end{flushleft}
\end{table}

\subsubsection{Comparing With Instruction Tuned CodeLLMs}

Table~\ref{tab:evaluation_instruct} presents the evaluation results comparing CodeLLMs fine-tuned with \tool{} to those trained via instruction tuning. CodeLLMs fine-tuned by \tool{} consistently outperform their instruction-tuned counterparts in both code correctness and efficiency.

Specifically, \tool{} achieves substantial improvements in code correctness from 5.79\% to 51.15\% in terms of pass@1. Similarly, the runtime efficiency of \tool{}-fine-tuned CodeLLMs surpasses instruction-tuned models by 9.92\% to 23.18\% in terms of GET score, highlighting the superior efficiency of \tool{} in generating correct and efficient code.

Interestingly, instruction tuning does not consistently yield improvements over the original CodeLLMs in either metric. Among the evaluated models, only Magicoder-S-DS-6.7B achieves a higher pass@1 compared to its original version, and only DeepseekCoder-6.7B-instruct and Magicoder-S-DS-6.7B show efficiency gains over their original versions (i.e., ECC > 0.5).
We attribute this inconsistency to the inherent limitations of instruction tuning, which heavily depends on diverse and representative training datasets. 
In this study, EffiBench provides a single ground-truth solution per task, which may not be sufficient to effectively guide CodeLLMs in generating both efficient and correct code. 
Furthermore, instruction tuning typically employs next-token prediction as its training objective, which focuses on predicting subsequent tokens based on the context rather than explicitly optimizing CodeLLMs for multi-objective and fine-grained goals. 
However, for the task of generating both correct and efficient code, the optimization goals encompass both code correctness and efficiency, necessitating an optimization process that extends beyond the capabilities of standard instruction tuning.
Conversely, \tool{} overcomes these challenges by utilizing a reinforcement learning-based fine-tuning method that 1) autonomously generates varied responses for each coding task to enable iterative training and 2) integrates multi-objective optimization directly into its reward system.

\subsubsection{Comparison with PIE}
We compare the performance of CodeLLMs fine-tuned by \tool{} with those fine-tuned by PIE~\cite{shypula2023learning} in terms of code correctness and efficiency. 
The evaluation results are shown in Table~\ref{tab:evaluation_pie}.

CodeLLMs fine-tuned by \tool{} consistently outperform PIE-tuned models in both code correctness and efficiency. 
In particular, \tool{} enhances code correctness over PIE by a margin ranging from 2.92\% to 14.41\% in terms of pass@1.
The runtime efficiency of code generated by \tool{}-fine-tuned CodeLLMs surpasses that of PIE-tuned models by 5.31\% to 11.45\% in terms of GET score.
The results suggest that \tool{} is more effective in optimizing code efficiency and correctness than PIE.

Furthermore, we observe that though PIE improves the efficiency of generated code for all CodeLLMs (i.e., all ECC>0.5), it does not consistently enhance code correctness. 
Specifically, CodeQwen1.5-7B-Chat and WaveCoder-Ultra-6.7B show lower pass@1 scores than the vanilla models that did not undergo fine-tuning.
This finding is aligned with existing studies~\cite{waghjale2024ecco} that highlight the trade-off between code efficiency and correctness in PIE. 
In contrast, \tool{} simultaneously enhances both code efficiency and correctness, overcoming these limitations. 

\begin{table}[]
    \caption{Correctness and Efficiency of Code Generated by CodeLLMs fine-tuned with PIE and \tool{}.}
    \begin{tabular}{@{}lllll@{}}
    \toprule
    Model Name                                   & Fine-Tune & ECC (\%) & GET  & Pass@1 (\%) \\ \midrule
    \multirow{2}{*}{CodeQwen1.5-7B-Chat}         &   PIE       & 62.3   & $1.13 \times 10^{-3}$ & 40.88       \\ \cmidrule(l){2-5} 
                                 &  \tool{}         & 72.3 & $1.07 \times 10^{-3}$ (\textcolor{red}{\textdownarrow 5.31\%}) & 46.77 (\textcolor{red}{\textuparrow 14.41\%})       \\ \cmidrule(l){1-5} 
    \multirow{2}{*}{DeepseekCoder-6.7B-instruct} & PIE          & 56.1   & $9.17\times 10^{-4}$ & 44.61       \\ \cmidrule(l){2-5} 
                                                 &  \tool{}          & 61.2 & $8.12 \times 10^{-4}$ (\textcolor{red}{\textdownarrow 11.45\%}) & 48.41 (\textcolor{red}{\textuparrow 8.51\%})       \\ \cmidrule(l){1-5} 
    \multirow{2}{*}{WaveCoder-Ultra-6.7B}        & PIE          & 59.7   & $1.65\times 10^{-3}$ & 42.41       \\ \cmidrule(l){2-5} 
                                                 &  \tool{}          & 65.4 & $1.55\times 10^{-3}$ (\textcolor{red}{\textdownarrow 6.06\%}) & 43.65 (\textcolor{red}{\textuparrow 2.92\%})       \\ \cmidrule(l){1-5} 
    \multirow{2}{*}{Magicoder-S-DS-6.7B}         & PIE          & 58.3   & $1.15\times 10^{-3}$ & 52.46       \\ \cmidrule(l){2-5} 
                                                 &  \tool{}          & 67.9 & $1.07\times 10^{-3}$ (\textcolor{red}{\textdownarrow 6.96\%}) & 56.35 (\textcolor{red}{\textuparrow 7.42\%})       \\ \bottomrule
\end{tabular}
\label{tab:evaluation_pie}

\end{table}

\begin{tcolorbox}[colback=blue!5!white, colframe=blue!75!black, title=Key Findings]
\noindent \textbf{Summary:} 
The efficiency and correctness of code generated by CodeLLMs fine-tuned by \tool{} are improved compared to the original CodeLLMs by 0.52\% to 10.86\% in terms of GET score and by 1.84\% to 14.51\% in terms of pass@1 for four state-of-the-art CodeLLMs. 
Furthermore, \tool{} outperforms instruction tuning and PIE in terms of code correctness and efficiency, achieving improvements of up to 51.15\% in pass@1 and 23.18\% in GET score compared to instruction tuning, and up to 14.41\% in pass@1 and 11.45\% in GET score compared to PIE.
\end{tcolorbox}

\subsection{RQ2: Impact of \tool{} Parameters on CodeLLMs Performance}

We conduct ablation studies to investigate the impact of \tool{} parameters on the performance of CodeLLMs fine-tuned by \tool{}.
We identify the following factors of \tool{} that may affect its performance: (1) the training epoch number and (2) the number of CodeLLM responses.  

\begin{figure}[ht]
    \centering
    \includegraphics[width=0.8\textwidth]{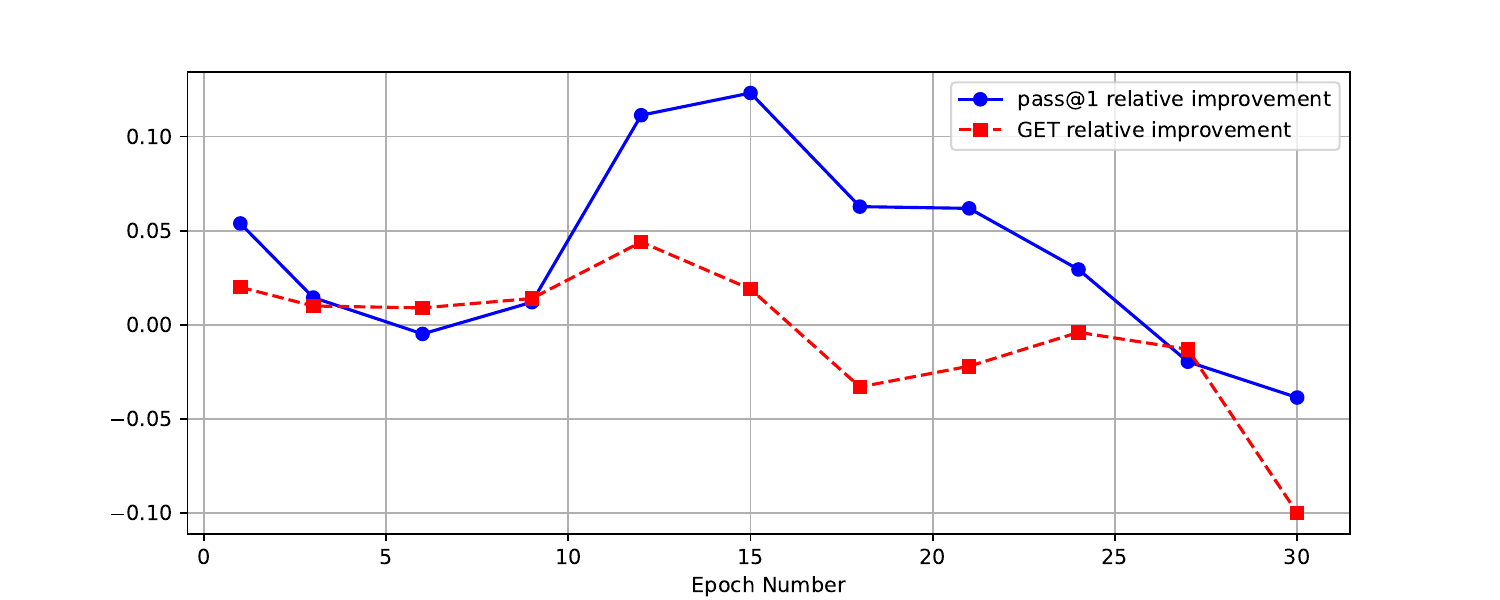}
    \caption{Impact of Training Epoch Numbers on \tool{} Performance. Pass@1 and GET relative improvements refer to the percentage of improvement in terms of Pass@1 and GET scores compared to the original CodeLLMs.}
    \label{fig:epoches}
\end{figure}

\subsubsection{Number of Training Epoches}
\textbf{Experiment Setting:}
To investigate the impact of the training epoch on the performance of \tool{}, 
we compare the performance of CodeQwen1.5-7B-Chat fine-tuned by \tool{} with different training epoch numbers ranging from 1 to 30.
We set the epoch interval to be 3.

\noindent \textbf{Results:}
The evaluation results are shown in Figure~\ref{fig:epoches}.
We observe that epoch number affects the performance of \tool{}.
Specifically, as the training epoch number increases from 1 to 18, the GET score of tuned CodeLLMs is consistently higher than that of the original CodeLLMs by 0.01 to 0.05.
Similarly, the Pass@1 score of tuned CodeLLMs is also higher than the original CodeLLMs by up to 11.52\% from 1 to 24 epoch except for the 6th epoch, which is 0.01\% lower than the original CodeLLMs.
The improvement variance may be due to the randomness of the RL training process.
However, we observe that the performance of tuned CodeLLMs in terms of GET and Pass@1
starts to decrease and performs worse than the original CodeLLMs consistently
 after 18 and 24 epochs, respectively.
We hypothesize that overfitting may occur when the number of training epochs is too large, which leads to a performance decrease of \tool{}.

\begin{figure}[h]
    \centering
    \includegraphics[width=0.8\textwidth]{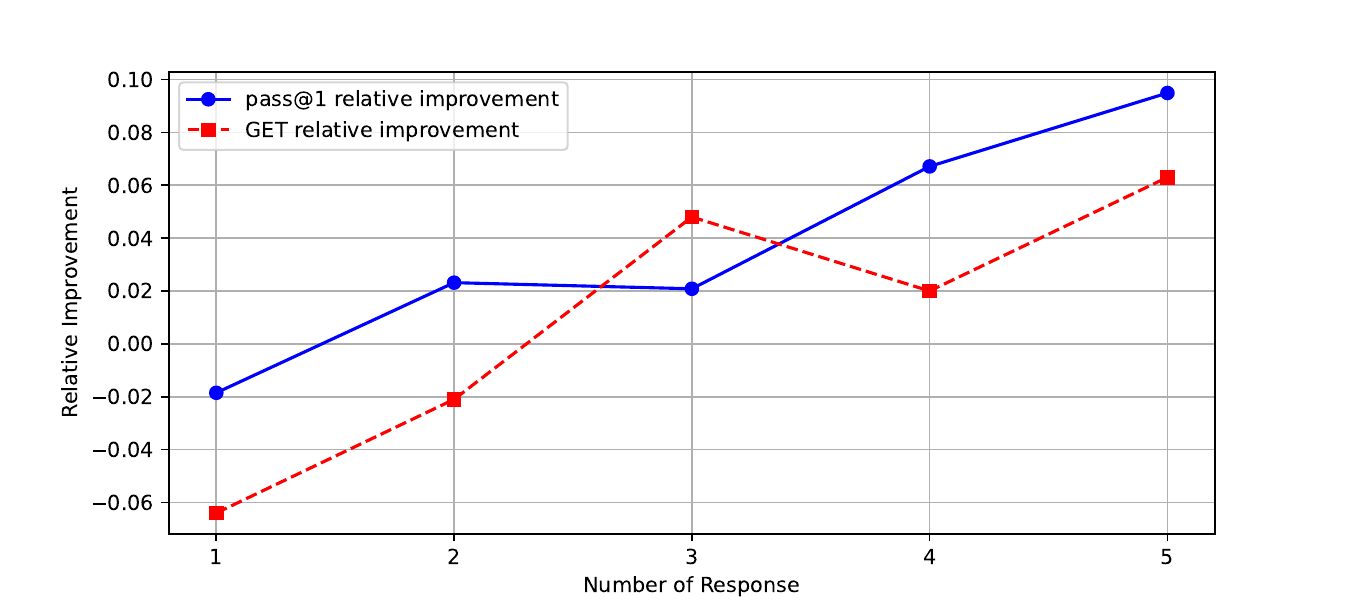}
    \caption{Impact of Number of Responses on \tool{} Performance. Pass@1 and GET relative improvements refer to the percentage of improvement in terms of Pass@1 and GET scores compared to the original CodeLLMs.}
    \label{fig:response}
\end{figure}

\subsubsection{Number of CodeLLM Response} 
\textbf{Experiment Setting:}
We investigate the impact of response number $N$ on the performance of \tool{}.
% Specifically, we analysis the impact of response number for each data in the training dataset.
A response number $N$ indicates that \tool{} generates N responses for each coding task in the training dataset, which leads the pairs of coding tasks and solutions that are used for RL training to be $N$ times the size of the original dataset. 
We conduct experiments using $N$ ranging from 1 to 5 for fine-tuning CodeQwen1.5-7B-Chat on \tool{}.

\noindent \textbf{Results:}
The evaluation results are shown in Figure~\ref{fig:response}.
We can observe that both GET and Pass@1 scores of tuned CodeLLMs generally increase as the response number increases from 1 to 5, except for the GET score when the response number is 4 and the pass@1 score when the response number is 3.
We attribute these slight declines to potential variability in RL training.
Specifically, both GET and Pass@1 scores achieve the highest improvement when the response number is 5, indicating that the performance of \tool{} is positively correlated with the response number and follows the trend of scaling law.
However, the training time cost of \tool{} is approximately linearly increasing with the response number, presenting a trade-off between the performance and training cost of \tool{}.

\section{Threats to Validity}

The threats to external validity concern the generalizability of our findings. To mitigate this threat, we evaluated \tool{} using four popular and state-of-the-art CodeLLMs on the Evalplus~\cite{evalplus} benchmark. However, we acknowledge that our findings may not generalize to CodeLLMs of other sizes or architectures. In the future, we plan to extend our study to a broader range of CodeLLMs to further validate the effectiveness of \tool{}.

The threats to internal validity of this study include the measurement of execution time and correctness of CodeLLM-generated code. 
To mitigate this threat, we run each code snippet with an adaptive repeated strategy to ensure the code snippets are executed multiple times and take the average execution time as the final result. Meanwhile, different from the previous benchmark EffiBench, which uses profiling tools to measure the execution time of code snippets, we follow the best practices in Python documentation for runtime benchmarking~\cite{docprofile} and use \textit{timeit} module. Furthermore, we take into account the library import overhead and exclude its time cost from the final result by leveraging the \texttt{timeit} module. On the other hand, to mitigate the threat of randomness in code correctness evaluation, we fix the temperature of LLM as 0 and $top_{p}$ as 1 to force LLMs to generate identical code snippets for each coding task.

The threats to construct validity mainly lie in the selection of evaluation metrics. To mitigate this threat, we identify the deficiencies of existing evaluation metrics and propose two new evaluation metrics to better measure the performance of CodeLLMs. 

% \section{Related Work}

\section{Conclusion}
In this paper, we propose \tool{}, a reinforcement learning-based fine-tuning framework for CodeLLMs, to improve CodeLLMs' performance in terms of both the correctness and efficiency of their generated code. 
\tool{} leverages execution feedback of the generated code to guide reinforcement learning-based fine-tuning of CodeLLMs, in which the reward is calculated based on the execution correctness and runtime efficiency of the generated code. 
We conduct experiments on four state-of-the-art CodeLLMs and evaluate the performance of CodeLLMs fine-tuned on \tool{} in terms of code correctness and efficiency.
The results show that CodeLLMs fine-tuned by \tool{} outperform the original CodeLLMs in terms of code correctness and efficiency by from 1.84\% to 14.51\% in terms of pass@1 and from 0.52\% to 10.86\% in terms of GET score.
In the future, we plan to extend our study to more CodeLLMs with different architectures, pre-training objectives, and parameter sizes.

%%
%% The acknowledgments section is defined using the "acks" environment
%% (and NOT an unnumbered section). This ensures the proper
%% identification of the section in the article metadata, and the
%% consistent spelling of the heading.
% \begin{acks}
% To Robert, for the bagels and explaining CMYK and color spaces.
% \end{acks}

%%
%% The next two lines define the bibliography style to be used, and
%% the bibliography file.
\bibliographystyle{ACM-Reference-Format}
\bibliography{sample-base}

%%
%% If your work has an appendix, this is the place to put it.
\appendix

\end{document}